%% file: main.tex
\documentclass[journal,onecolumn,tikz]{IEEEtran}
\IEEEoverridecommandlockouts
\usepackage{cite}
\usepackage{amsmath,amssymb,amsfonts}
\usepackage{algorithmicx}
\usepackage{algorithm}
\usepackage{algpseudocode}
\usepackage{graphicx}
\usepackage{textcomp}
\usepackage{xcolor}
\usepackage{soul}
\usepackage[hidelinks]{hyperref}
\usepackage{cleveref}
\usepackage{ragged2e}
\usepackage{xcolor}
\usepackage{tikz}
\usepackage{comment}
\usepackage[margin=1in]{geometry}
\usepackage{xspace}
\usepackage{enumitem}

\makeatletter
\renewenvironment{abstract}{%
  \normalfont
  \@IEEEabskeysecsize\bfseries\textit{\abstractname--}%
  \@IEEEgobbleleadPARNLSP
}{%
  \par\ifCLASSOPTIONconference\vspace{0ex}\else\vspace{1.34ex}\fi
  \normalfont\normalsize
}
\makeatother

\newcommand{\iic}{$\text{I}^2\text{C}$\xspace}

\newcommand{\sys}{\textsc{Duet}\xspace}
\newcommand{\baseline}{\textsc{Baseline}\xspace}

\def\BibTeX{{\rm B\kern-.05em{\sc i\kern-.025em b}\kern-.08em
    T\kern-.1667em\lower.7ex\hbox{E}\kern-.125emX}}
\begin{document}
\usetikzlibrary{arrows.meta, positioning}
\tikzset{
  box/.style = {rectangle, draw, rounded corners, minimum width=2.8cm, minimum height=1cm, align=center},
  arrow/.style = {-{Latex[length=3mm,width=2mm]}, thick}
}
\title{\textsc{Duet}: Agentic Design Understanding via\\Experimentation and Testing}

\author{\IEEEauthorblockN{Gus Henry Smith$^1$*},
\IEEEauthorblockN{Sandesh Adhikary$^2$}, 
Vineet Thumuluri$^2$,
Karthik Suresh$^2$,
Vivek Pandit$^2$,\\
Kartik Hegde$^2$,
Hamid Shojaei$^2$,
Chandra Bhagavatula$^2$
\\

$^1$Southmountain Research, $^2$ChipStack**\\
\footnote{*work done while at ChipStack; **work done before ChipStack acquisition by Cadence Design Systems}
\texttt{gus@southmountain.ai}, \texttt{\{sandesha,vineett,sureshk,vivekp, kartikv,hamids,bchandra\}@cadence.com}
}

\maketitle
\thispagestyle{empty}
\pagestyle{empty}


\begin{abstract}
\input{text/abstract}
\end{abstract}


\input{text/intro}
\input{text/duet}
\input{text/evaluation}

\input{text/related-work}

\input{text/conclusion}

\bibliographystyle{unsrt}  
\bibliography{text/refs}           
\end{document}

%% file: text/abstract.tex
AI agents
  powered by large language models (LLMs)
  are being used to solve increasingly
  complex software engineering challenges,
  but struggle
  with hardware design tasks.
Register Transfer Level (RTL) 
  code 
  presents a unique challenge for LLMs, as it
  encodes
  complex, dynamic, time-evolving 
  behaviors
  using the low-level language features
  of SystemVerilog.
LLMs struggle 
  to infer these complex behaviors
  from the syntax of RTL alone,
  which limits their ability to complete
  all downstream tasks
  like 
  code completion, documentation, or verification.
In response to this issue,
  we present \sys:
  a general methodology for developing  
  \underline{D}esign \underline{U}nderstanding
  via
  \underline{E}xperimentation and \underline{T}esting.
\sys{} mimics how hardware design experts
  develop an understanding of complex designs:
  not just via a one-off readthrough 
  of the RTL,
  but via iterative experimentation
  using a number of tools.
\sys{} iteratively generates hypotheses, 
  tests them with EDA tools 
  (e.g., simulation, waveform inspection, and formal verification), 
  and integrates the results to build a bottom-up understanding of the design.
In our evaluations, we show that \sys{}
  improves AI agent performance on 
  formal verification,
  when compared to a baseline flow
  without experimentation.

%% file: text/intro.tex
\section{Introduction}
\label{sec:intro}

\textit{AI agents}
  are quickly taking over many tasks 
  in software engineering and beyond. Powered by large language models (LLMs), AI agents can be deployed as powerful autonomous software workflows.
An AI agent takes
  a text prompt as input, and 
  is generally also equipped
  with tools it can call
  (e.g., Python functions or command-line utilities
  to read and write files, execute computations,
  or run installed tools).
The agent then iteratively queries the 
  LLM (starting with the prompt)
  to get the next action.
When the LLM requests an action,
  the function or command-line utility is called
  and the results are sent back to the LLM.
Eventually, the LLM
  decides to stop (or is stopped externally),
  and sends a final result.
Throughout the process, the agent
  may have also created
  a number of artifacts,
  such as files in the filesystem.
Using this general structure,
  AI agents have been able to replicate
  many human tasks.

While AI agents
  have shown human-level performance
  in software engineering tasks,
  they continue to struggle with 
  similar tasks in hardware design.
Among many reasons, we hypothesize that this is because
  the underlying LLMs powering AI-assisted tools
  struggle with 
  Register Transfer Level (RTL) code.
RTL
  is inherently complex, 
  often capturing
  dynamic, time-evolving 
  behaviors 
  using the low-level,
  heavily implicit
  syntax of languages like SystemVerilog.
This is in stark contrast to software languages
  like Python or Java
  which contain more structure
  in the syntax of the code itself;
  for example, sequential lines 
  in imperative programming languages
  generally correspond to instructions which will execute over time,
  and function names can be used to understand
  when control flow jumps across files in the codebase.
On the other hand, RTL
  does not have such structure.
For example, sequential states in a finite state machine (FSM)
  might be separated by tens or hundreds of lines
  in a \texttt{case} statement,
  and the control flow between these states may be much
  less explicit.
Thus, when LLMs are given only the RTL,
  they struggle to
  understand the deeper behaviors of hardware designs
  and build
  design understanding.
As a result,
  AI agents perform worse on
  downstream hardware tasks
  like verification or debugging.

However, hardware designers themselves
  do not build their understanding of a design
  simply by reading the RTL.
Instead, the process is more dynamic;
  designers will use tools like simulation,
  waveform viewers,
  and formal tools
  to understand the design.
Even if the designer builds 
  their understanding of the design
  purely from documentation,
  that documentation contains descriptions of the dynamic behavior 
  of the design such as timing diagrams and waveforms.

  
  We present \sys:
  a general methodology for developing  
  \underline{D}esign \underline{U}nderstanding
  via
  \underline{E}xperimentation and \underline{T}esting.
\sys{} that mimics how hardware design experts
  develop an understanding of complex designs:
  not just through reading the RTL,
  but via iterative experimentation.
  \sys{} enables AI agents to generate hypotheses about a design, and equips it with tools to test and confirm or refine these hypotheses.

Let's consider a simple example.
Imagine we task an AI agent with 
  describing a design
  with a reasonably complex finite state machine---%
  for example, an implementation of \iic. Specifically, let’s consider how an AI agent describes the \iic TX path clock stretching feature when it is given access to the RTL, but is not given any experimental tools. Below, in the left column of~\cref{fig:duet-outputs}, we present a 
  snippet of the output produced by such an agent:
\input{text/eval-figure}

\noindent
As we will soon see, this description
  is missing a key detail.
Imagine we use this description 
  as a starting point
  for a downstream task,
  such as
  having an AI agent
  verify the TX path clock stretching feature.
The agent will try and fail
  to verify the feature
  based on the description alone
  due to a lack of detail.

Now imagine we use \sys to enhance
  the process
  and produce deeper design understanding by giving the AI agent 
  access to experimentation tools
  and prompting it with experimentation
  strategies
  when it generates its description.
The agent decides to test its initial description
  using a simulation experiment.
It generates a simulation testbench
  and calls the provided simulation tool,
  resulting in the log shown in the middle
  column
  above.
These logs are sent back to the agent.
As we can see from the log,
  the agent encodes its hypotheses
  in the testbench itself;
  in this case,
  the agent's hypothesis
  (that its inputs would produce
    TX clock stretching behavior)
  proved to be false.
The agent analyzes this result, as shown in the second snippet in the middle column. The agent uses the analysis to fix the testbench
  in later iterations, and successfully exercises the feature (not shown).
This results in the improved feature description
  in the right column of~\cref{fig:duet-outputs}

This example demonstrates
  the key insight
  of this paper:
    despite the complexity of RTL,
    AI agents can learn
    deep design behaviors
    from
    simple trial-and-error experimentation.

\sys is a very general tactic that can be applied
  in any AI workflow operating over code
  to which experimentation tools can be applied.
For example,
  \sys could be used in documentation generation
  to produce deeper insights
  into specific features of the design,
  or it could be used in formal verification
  to examine counterexamples to properties which the agent
  is struggling to prove.
Though it is out of the scope of this paper,
  it is likely even useful outside of hardware and EDA 
  applications.

In our evaluation, we show that utilizing
  \sys{} 
  leads to 
  better performance on
  AI-assisted EDA tasks---%
  in the case of our evaluation,
  formal verification.
We compare a baseline flow,
  which uses an AI agent without
  access to experimentation tools,
  to a \sys-enabled flow.
Using experimentation,
  the \sys-enabled flow is better able
  to understand the design and
  debug verification failures.

The rest of this paper proceeds as follows.
In \cref{sec:implementation},
  we describe the implementation of \sys{}.
In \cref{sec:evaluation}, we
  evaluate \sys against a baseline AI workflow
  on
  the task
  of formal verification.
In \cref{sec:related-work},
  we present related work.
In \cref{sec:conclusion} we conclude.

%% file: text/eval-figure.tex
\begin{figure}[h]
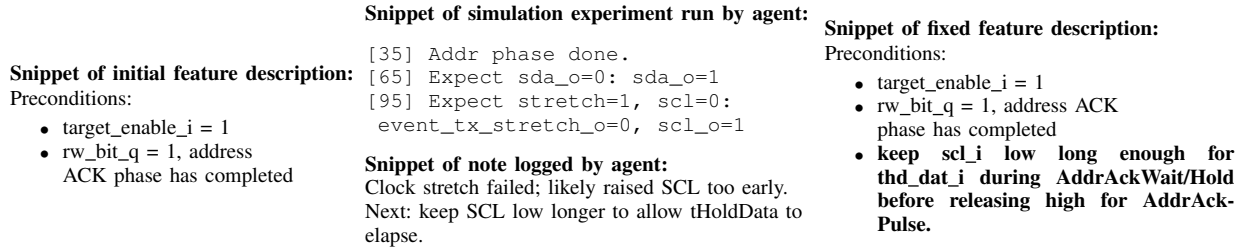


\footnotesize

\begin{minipage}{.28\linewidth}
\begin{FlushLeft}
\textbf{Snippet of initial feature description:}
\end{FlushLeft}
\vspace{-2mm}
Preconditions:
\begin{itemize}
    \item target\_enable\_i = 1
\item rw\_bit\_q = 1, address \\ ACK
  phase has completed
\end{itemize}
\end{minipage} 
\begin{minipage}{.37\linewidth}

\begin{FlushLeft}
\textbf{Snippet of simulation experiment run by agent:}
\end{FlushLeft}
\begin{verbatim}
[35] Addr phase done.
[65] Expect sda_o=0: sda_o=1
[95] Expect stretch=1, scl=0:
 event_tx_stretch_o=0, scl_o=1
\end{verbatim}

\begin{FlushLeft}
\textbf{Snippet of note logged by agent:}

Clock stretch failed; 
likely raised SCL too early. 
Next: keep SCL low longer 
to allow tHoldData to elapse.
\end{FlushLeft}

\end{minipage}\begin{minipage}{.33\linewidth}

\begin{FlushLeft}
\textbf{Snippet of fixed feature description:}
\end{FlushLeft}
\vspace{-2mm}
Preconditions:
\begin{itemize}
\item target\_enable\_i = 1
\item rw\_bit\_q = 1, address ACK \\
  phase has completed
    \item 
\textbf{keep scl\_i low long enough 
  for thd\_dat\_i during 
  AddrAckWait/Hold before 
  releasing high for 
  AddrAckPulse.}
\end{itemize}

\end{minipage}
\caption{Comparing illustrative design descriptions produced by a baseline LLM flow (left) and the same flow enhanced with \sys (right), along with snippets of \sys's experimentation and failure analysis (center).}
\label{fig:duet-outputs}
\end{figure}










%% file: text/duet.tex
\section{\sys}
\label{sec:implementation}

In this section we provide details
  on \sys itself.
We present one potential
  implementation of \sys,
  describe the 
  experimentation tools we have used,
  and cover pitfalls in implementation.

\subsection{\sys as a Subroutine}
\label{sec:impl:duet-as-subroutine}

\newcommand{\doExperiment}{DoExperimentation}

While \sys is an intentionally generic methodology
  with many possible implementations,
  in this section we provide the implementation
  of a ``canonical'' form
  of \sys:
  the 
  \textsc{\doExperiment} procedure.
\textsc{\doExperiment}
  implements the heart of \sys:
  the experimentation loop.
Given the design, 
  an experiment description (``expt''), 
  and a set of tools,
  \textsc{\doExperiment} carries out the experiment
  using the provided set of tools:

\clearpage

\begin{algorithmic}
\Procedure{\doExperiment}{design, expt, tools}
    \State $\text{tools} \gets \text{tools} + [ \textsc{EndExperimentation}] $
    \State $\text{messages} \gets [
    \text{``given the following design:''} + \text{design} +
    \text{``run the following experiment:''} + \text{expt}  ]$ 
    \Loop \Comment{Core experimentation loop: each iteration is a single experiment}
        \State $\text{response} \gets \Call{LLM}{\text{messages}, \text{tools}}$; $\text{output} \gets \text{run tool in response}$; $\text{messages} \gets \text{messages} + [\text{response},  \text{output}]$
        \State \algorithmicif\ called \textsc{EndExperimentation} \algorithmicthen\ break from loop
    \EndLoop
    \State $\text{report} \gets  \Call{LLM}{\text{messages} + [ \text{``generate an experiment report using the previous context''}]}$; \Return{report}
\EndProcedure
\end{algorithmic}

The AI agent has full control over
  what experiment it runs in each iteration
  and when to end experimentation.
At the end of experimentation, 
  the agent
  produces an experiment report
  describing the hypotheses made,
  the tests run,
  and the conclusions drawn from the results.
Note that modern AI agent programming paradigms
  (such as the OpenAI Agents SDK \cite{openai_agents_python})
  make it easy to implement this experimentation loop;
  you simply need to provide the prompt and tools,
  and the loop will be handled internally.

\textsc{\doExperiment} 
  is a flexible base implementation
  of \sys in that it takes 
  a text description of the larger experiment to be performed as input
  and returns a text report
  of the results as output.
Starting from this base,
  a number of features could be added,
  e.g.~taking other inputs 
  (such as error logs from formal tools
    or PPA/timing results from backend tools)
  or producing richer outputs
  (such as working testbenches demonstrating
    the discoveries made during experimentation).




\subsection{Tools}
\label{sec:impl:tools}

\sys{} can be flexibly extended to use many tools.
Thus far, we have tried
  simulators,
  formal verification tools,
  waveform viewers,
  and
  tools for design analysis/signal extraction.
In addition,
  we have also used tools
  which are themselves agents---%
  see the counterexample replication tool below.
What follows is a description of the tools 
  we have used.

\paragraph*{Simulation}
The simulation tool allows the agent to run
  a testbench in simulation
  and see the printed logs.
The agent also sees any errors
  produced by the simulator,
  in the case of a malformed testbench. While the agent is free
  to write whatever testbench it would like,
  our prompts include various tips about
  testbench structure and
  common SystemVerilog pitfalls.
For example,
  it is useful to instruct the agent
  not to use assertions,
  which may fail and terminate the testbench early,
  but instead to use copious print statements
  to produce rich logs.
  
In our evaluations, we noticed that AI agents seem to rely most heavily on
  simulation-based experimentation.
There are a number of possible explanations.
First,
  even for hardware designers,
  simulation is straightforward to understand
  when compared to other tools.
Second,
  there is likely a bias towards
  simulation
  in the training data of LLMs,
  as simulation-based tests
  are presumably far more common
  in open-source repositories
  than formal testbenches.

\paragraph*{Waveform viewer}
Waveform viewing tools
  can be a useful supplement to simulation-based
  experimentation.
Waveforms produced by simulation experiments
  are generally too large 
  to include in the response to the LLM;
  however, given access to a tool like 
  \texttt{vcdcat},
  agents can decide which portions of the waveform
  to view.
  
\paragraph*{Formal}
Formal verification tools such as Jasper
  give the agent the ability to prove properties
  or check whether certain states are reachable
  using coverpoints.
As we show in \cref{sec:evaluation},
  the agent generally does not use
  formal verification for experimentation,
  though we suspect that with better prompt engineering and few-shot-examples, the agent could be tuned to utilize formal verification more heavily.
  
\paragraph*{Counterexample replication}
The counterexample replication tool
  is itself an AI agent,
  which takes a 
  formal property
  and a counterexample found by 
  a formal verification tool
  and converts the counterexample
  into a SystemVerilog simulation testbench
  which replicates the counterexample.
Each call to the counterexample replication tool
  generally involves 1--5 calls
  to the simulation tool,
  in which the agent attempts to produce
  a working replication.
As we see in \cref{sec:evaluation},
  this proves to be a very valuable tool
  as
  the trial-and-error process
  of counterexample replication
  seems to produce deeper design understanding.

\paragraph*{Signal extraction and design analysis}
We have also experimented with using
  tools for signal extraction
  and design analysis.
For example, the Yosys~\cite{wolf2013yosys}
  open-source synthesis tool
  provides a number of useful utilities
  such as FSM detectors.
These types of tools can provide useful
  high-level information
  in a deterministic manner.
However, similar to the formal tool above,
  they require careful prompting
  so the agent knows how and when
  to use them.

\subsection{Pitfalls}

We have seen a few pitfalls
  when implementing an experimentation agent.
As with all agents,
  experimentation agents are good
  at finding ways to ``cheat''.
For example, if not properly constrained,
  the counterexample replication tool
  will simply force internal signals of the design
  such that the counterexample is trivially replicated,
  rather than replicating it using the module inputs.
Strict prompting fixes this issue.
The experimentation agent
  can also be too lax
  when gathering evidence for a claim,
  or in the worst case,
  can even hallucinate results;
  simply asking the agent to thoroughly cite 
  experiment logs
  for each of its claims
  can mitigate this issue.
All of these observations
  are a reminder that it is all too easy
  to think of modern LLM-based AI agents
  as ``human-like'' intelligence.
While they are incredibly powerful
  and useful,
  they also fail in ways
  most human experts 
  never would.

%% file: text/evaluation.tex
\section{Evaluation}
\label{sec:evaluation}

To evaluate \sys{},
  we integrate it
  into one of our existing AI-assisted workflows.
Our AI-assisted workflows
  tackle various EDA tasks specifically
  in the hardware verification space;
  in this evaluation, we will focus primarily
  on our formal verification flow.

\subsection{Experimental Setup}

In this subsection,
  we describe our two points of comparison:
  our baseline AI-assisted agentic formal flow (\baseline{})
  and our \sys-enabled flow (\sys).

\vspace{3mm}
\paragraph{\baseline{}}
Our baseline AI-assisted formal verification
  workflow
  works as follows.
Note that this description is high-level
  as proprietary details are not included.
\begin{enumerate}
    \item The AI agent summarizes the design by observing the RTL.
    \item Based on the design summary, the AI agent develops a verification plan
      by enumerating \textit{properties} to verify.
    \item For each property:
    \begin{enumerate}[label=\roman*.]
        \item The agent generates a formal testbench
          for the property.
        \item The agent calls a tool
          which runs the formal testbench
          through Jasper.
        \item Viewing the report from Jasper,
          the agent returns to step (i) and iterates on the formal testbench
          until every assertion passes
          or until it hits its iteration limit.
        \item The agent returns the final testbench
          (passing or non-passing).
    \end{enumerate}
    \item Verification terminates after 
      the agent has processed each property in the verification plan.
\end{enumerate}

The baseline formal flow outputs a set of testbenches verifying the properties set out in the verification plan.

\vspace{3mm}
\paragraph{\sys}
Our \sys-enabled workflow adapts the baseline workflow with a few key modifications:
\begin{itemize}
    \item 
      The agent's prompt
        instructs the agent
        to use \textit{experimentation} when debugging
        formal testbench failure.
      These instructions include example scenarios
        of when to use experimentation:
        for example, 
        to reproduce counterexamples reported by Jasper
        or to attempt verification on sub-properties.
    \item The agent is given access to three
      experimentation tools (a subset of the tools described in \cref{sec:impl:tools}):
    \begin{itemize}
        \item \textbf{Formal tool:}
          runs a formal verification testbench
            with Jasper
            and returns the proof
            status of asserts, coverpoints, and assumes.
        \item \textbf{Simulation tool:}
          runs a simulation testbench
          with Verilator
          and returns the logs to the agent.
        \item \textbf{Counterexample replication tool:}
          a special, more directed case of the simulation tool which
            uses simulation via Verilator
            to reproduce a Jasper counterexample.
    \end{itemize}
      
\end{itemize}
\vspace{3mm}
\paragraph{Running the workflows}
We run the two workflows 
  on the same design: a simple round-robin arbiter.
We use the same design summary
  and verification plan
  (steps 1 and 2 of \baseline{})
  for both flows,
  to ensure we can compare the outputs
  property-by-property.
We arbitrarily limit the 
  verification plan to
  a total of
  ten properties.
Each flow is only given a single attempt
  to prove each property.
We use GPT-5.0 for all LLM queries.
We use Jasper as our formal verification tool
  and Verilator as our simulator.

\subsection{Results and Analysis}

\begin{table}[]
    \centering
    \begin{tabular}{c|c|c|c||c|c|c}
         Property & \baseline{} & \sys{} & Improved & \# \sys sim expts & \# \sys repl expts & \# \sys formal expts \\
         \hline
1& unproven& proven & \checkmark & 0&7&0\\
2& unproven & unproven, refined & $\sim$ & 0&5&0\\
3&proven & proven & & 1 &4&0 \\
4&proven&proven & & 3&1&1\\
5&unproven & unproven & & 1&1&0 \\
6&unproven & proven & \checkmark &1&1&0 \\
7&vacuous & proven  & \checkmark & 1&1&0\\
8&unproven&unproven & & 0&1&0\\
9&proven&proven & & 1&1&0\\
10& unproven&unproven & & 3&1&1 \\
    \end{tabular}
    \vspace{1mm}
    \caption{\baseline{} vs.~\sys performance on the task of formally verifying properties of a design.}
    \label{tab:duetvsbaseline}
\end{table}

\Cref{tab:duetvsbaseline} shows
  the results
  of using both flows
  to verify the same 10 properties of the design.
We can see that the \sys-enhanced flow
  doubles the number of proven properties
  to six, over the three
  proven in the baseline.

In \Cref{tab:duetvsbaseline}, we classify a property as having been improved if \sys was able to prove a previously unproven or vacuous assertion in the \baseline{}. Two of the improved cases (1 and 6)
  were straightforward
  instances of
  the baseline flow failing to prove a property,
  while the \sys-enhanced flow
  successfully proved the property.
Perhaps more interesting
  were the other two cases.
Property 7 was proven but vacuous in the baseline.
However, via experimentation, the \sys-enabled flow was 
  able to find a non-vacuous
  refinement of the property
  and prove it.
Property 2 was unproven by the baseline,
  but the \sys-enabled flow successfully \textit{refined} it.
While not successfully proving the property,
  experimentation still uncovered a path towards progress (hence the $\sim$).

The last three columns of \cref{tab:duetvsbaseline}
  show the number of calls to each of the experimentation tools.
Recall that each call to the counterexample replication tool
  (indicated by the ``\# repl expts'' column) 
  actually represents multiple calls (generally 1-5)
  to the simulation experimentation tool.

The counterexample replication tool
  proved to be the most useful
  when attempting to make progress on formal verification.
Every improved property used the counterexample replication tool.
In the extreme case of property 1,
  the agent replicated seven separate counterexamples
  using the tool.
It is perhaps unsurprising that this tool proved
  the most useful;
  understanding counterexamples is core to the process
  of formal verification.

The simulation tool was generally useful as well,
  though not used on all improved properties.
Our general observation is that simulation
  is favored over formal in a way that suggests
  the LLM better understands how to craft
  SystemVerilog simulation testbenches,
  vs.~formal testbenches. The formal tool was rarely used---only in two cases.
There are 
  numerous other tools
  we didn't include in this evaluation
  which are treated largely the same
  by the agent---i.e., mostly ignored.
We suspect this could be fixed
  using better prompting,
  including clear examples
  of when each tool might be useful.
Additionally, 
  the ``formal experimentation tool'' is
  functionally identical
  to the tool which the agent uses to invoke Jasper
  on its output testbench,
  though the name is different
  and the prompt contains different instructions
  for using the tool.
It may be the case that 
  the agent gets confused by the presence of both tools.
Indeed,
  when the agent iterates on the final testbench
  by writing a testbench, calling the formal tool,
  and viewing the logs,
  it is simply a form of experimentation.

The evaluation also
  highlights where \sys{}
  can be improved.
For example, despite careful prompting,
  there were still some cases
  where the agent found ways
  to ``cheat''---%
  for example, replicating
  counterexamples by simply
  forcing the outputs
  to the expected values,
  instead of only manipulating
  the inputs.
Interestingly, though,
  some of these cases still seemed
  to help the agent with
  its understanding of the design,
  suggesting that simply reformatting
  the counterexamples
  into a more comprehensible format
  is itself useful.

\paragraph*{Case Study: Property 1}
To peek inside the \sys{} flow,
  we will look at flow artifacts for Property 1.
Both the baseline and \sys{}-enabled flows begin
  by an RTL-only analysis of the design,
  in which the agent reads the design files
  and writes out a design summary,
  which we show in truncated form below:

\begin{center}
\begin{verbatim}
Design summary (as implemented):
- Module and parameters
  - Top: arbiter
  - Parameters:
    - NUM_PORTS: number of request/grant ports ...
  - Ports:
    - clk: clock, rising-edge
    - rst: synchronous reset (active high) ...
- Core state and combinational signals ...
- Sequential behavior per clock edge ...
- What to verify (suggested properties) ...
\end{verbatim} 
\end{center}

\noindent
As this summary comes only from
  viewing the RTL,
  it may be incorrect.
It is on the verifier agent
  to realize this
  and refine its properties accordingly.

We then begin the actual process of verification with a prompt:

\begin{verbatim}
Property 001: grant_subset_of_request
Goal: Prove that in every cycle, grant is a subset of request:
  (grant & ~request) == 0.
...
\end{verbatim}

\noindent
As we will soon see, this property isn't correct as stated;
  however, the agent will be able to discover this 
  and refine the property 
  via experimentation.
After some reasoning,
  the agent produces its first attempt at a
  formal testbench.
We see this behavior often---%
  even when experimentation tools are available,
  the agent will often attempt to achieve the goal
  without them first.
This behavior can be controlled via prompting,
  but we have found it more beneficial to 
  focus on ensuring that the agent uses the experimentation tools
  if and when its initial attempts fail.

The following is a truncated version
  of its initial attempt at a formal
  testbench:

\begin{verbatim}
module arbiter_grant_subset_sva #(...) (...);
  property p_grant_subset;
    @(posedge clk) (grant & ~request) == '0;
  endproperty
  assert property (p_grant_subset);
endmodule
\end{verbatim}

\noindent
When the agent calls Jasper on
  this initial attempt
  at a formal testbench,
  Jasper produces counterexamples.
The agent then uses the counterexample replication tool.
Recall that the counterexample replication tool
  is itself an agent
  which uses multiple calls to the simulation tool
  to replicate a given counterexample.
Below we show the truncated output
  of the counterexample replication tool:

\begin{verbatim}
module tb_arbiter_ce;
  ...
  arbiter #(...) dut (...);
  initial clk = 1'b0; always #5 clk = ~clk;
  int cyc; initial cyc = 0; always @(posedge clk) cyc <= cyc + 1;

  // Basic stimulus to steer token to index 4 and produce a grant on bit 4
  initial begin ... end

  // Create and check the counterexample right when grant[4] goes high
  always @(posedge clk) begin
    if (cyc == 2) begin
      rst     <= 1'b1; request <= 6'h22; #1;
      assert (grant[4] == 1'b1 && request[4] == 0);
    end
  end
endmodule
\end{verbatim}

\noindent
Though the agent
  doesn't directly use the output
  of the counterexample replication tool,
  it seems that the
  process of replication 
  gives the agent a deeper understanding
  of the design.
In this case, it shows the agent
  that the timing behavior of the given property
  is more complex than originally proposed,
  and helps the agent refine the property.
Finally, the agent is able to produce a formal testbench
  which passes:

\begin{verbatim}
module arbiter_grant_subset_sva #(...) (...);
  property p_grant_subset_prev_req;
    @(posedge clk)
      1'b1 |-> ##1 ((grant & ~$past(request)) == '0);
  endproperty
  assert property (p_grant_subset_prev_req); 
endmodule
\end{verbatim}

Compared to the \sys{}-enabled flow,
  the baseline flow
  is unable to discover the exact timing
  of the property.
While it iterates a number of times
  on the formal testbench,
  it fails to pull deeper meaning
  from the counterexamples returned
  by Jasper.

%% file: text/related-work.tex
\section{Related Work}
\label{sec:related-work}

\subsection{LLMs for Software Test Generation}
Large Language Models (LLMs) have shown remarkable progress in software test generation. Most existing frameworks for LLM-based software-testing utilize a multi-step pipeline comprised of code-understanding, followed by test-generation, and finally a correction loop where testbenches are fixed based on compiler-feedback. For instance, frameworks including TestSpark \cite{Sapozhnikov2024TestSparkII} for Java and TestPilot \cite{schafer2024testpilot} for Javascript/Typescript follow such a generation-validation-repair pipeline. ChatUnitTest \cite{Chen2023ChatUniTestAF} additionally utilizes an adaptive focal context generation module for modular code-understanding. Similarly, ChatTester \cite{Yuan2023NoMM} also leverages ChatGPT to first understand the focal method, and was able to improve both the compilability and quality of generated assertions compared to direct application of ChatGPT. In the TestART framework \cite{Gu2024TestARTIL}, the final feedback loop also includes a coverage report -- prompting the model to update the testbench to also fill coverage holes. Other methods such as HITS~\cite{wang2024hits} have improved coverage through method slicing, achieving higher branch and line coverage on complex Java benchmarks compared to EvoSuite \cite{Fraser2011EvoSuiteAT}. An alternative approach to coverage maximization utilized by methods such as MuTAP \cite{Dakhel2023EffectiveTG} leverages mutations to maximize coverage by injecting bugs into the code and prompting the LLM to update testbenches to cover any uncaught bugs. There has also been recent work that uses fine-tuning and reinforcement learning to improve LLM's reasoning abilities for better software testing. For example, \cite{Steenhoek2023ReinforcementLF} fine-tune models to minimize test-smells in unit-tests, and \cite{Wang2025CoEvolvingLC} co-evolve the training of code-generation and unit-test-generation to improve performance for both tasks. 

Agentic frameworks such as \textit{ChatDev}~\cite{qian2023chatdev}, and \textit{SWE-Agent}~\cite{yang2024sweagent} introduced iterative LLM–tool loops, where models perform reasoning, invoke compilers or linters, and refine outputs based on tool feedback. Recent extensions to verification, like \textit{Prompt. Verify. Repeat.}~\cite{hassan2025pvr}, adopt similar human-in-the-loop workflows for hardware, coupling LLM reasoning with EDA tools (simulators, formal analyzers) to validate and correct generated artifacts. 
These “centaur” systems demonstrate that tool-driven iteration can improve reliability but also highlight a lack of structured experimentation frameworks—LLMs must reason not only over text but over the dynamic state of design execution.

The successful application of LLMs in software testing -- particularly agentic LLM systems -- demonstrates the promise of the same for hardware verification. However, this domain presents some unique challenges that warrant domain-specific solutions. As discussed earlier, RTL code lacks the explicit structural cues that software languages provide, making it difficult for LLMs to understand deeper hardware behaviors and resulting in worse performance on verification tasks. Such deeper design-understanding is possible through \sys's iterative experimentation.

\subsection{LLM Agents for Hardware Verification}
Recent work has begun exploring LLMs for hardware verification, particularly for formal property generation and testbench synthesis. 
\textit{AssertLLM}~\cite{xie2025assertllm} converts unstructured hardware specifications (text and waveform diagrams) into SystemVerilog Assertions (SVAs) using a two-stage LLM pipeline. 
It achieves higher syntactic accuracy than GPT-3.5 or GPT-4 on handcrafted benchmarks but still fails to guarantee semantic correctness, often missing temporal operators or misidentifying relevant signals. 
Similarly, \textit{AssertionBench}~\cite{pulavarthi2025assertionbench} provides a standardized benchmark for evaluating LLMs on real-world SVA and PSL properties, showing that none of
the LLM models evaluated could generate valid assertions an
average of no more than 44\% accuracy whereas
up to 63\% generated assertions produced CEX and
on average up to 33\% of generated assertions were
syntactically wrong. A crucial bottleneck is obtaining a precise understanding of the design from RTL, which is made challenging by the limited context-windows of LLMs. Existing work tends to circumvent this by splitting designs into hierarchical modules for independent analysis. \cite{Qayyum2024LateBR} proposes incremental proof-generation starting from the basic DUT module and expanding to additional modules. Similarly \cite{Hassan2024LLMGuidedFV}, use LLMs to hierarchically parse the DUT to obtain a formal model and invariants, and employ mutation testing to validate the invariants. Following a similar intuition, \sys also breaks down the task of design-understanding into smaller sub-tasks. But going beyond simply parsing the DUT into smaller modules, \sys allows breaking down the design in terms of functionalities, and builds a deep understanding of each functionality through verified outputs from deep experimentation with tools used by real hardware verification engineers.

In simulation-based verification, \textit{AutoBench}~\cite{qiu2024autobench} demonstrated that LLMs can generate self-checking testbenches from RTL module descriptions using a hybrid prompt-template framework. The model achieved a 57\% improvement in pass rate over baseline LLM models, but relied heavily on handcrafted scaffolding and lacked temporal reasoning. \cite{Mali2024ChIRAAGCI} extract a structured representation of design specifications, which are used to prompt an LLM to obtain a set of assertions. These assertions are then refined by iteratively prompting the LLM to fix issues identified by a simulator. LAAG-RV \cite{Maddala2024LAAGRVLA} follows a similar protocol, but additionally utilizes a `signal-synchronization' step to refine assertions by prompting an LLM to align the assertions with the design's RTL code.

In addition to using LLMs for verification, various works have also utilized LLMs to generate RTL code \cite{Thakur2023VeriGenAL, Ho2024VerilogCoderAV, Huang2024TowardsLV, Thakur2022BenchmarkingLL,Thakur2023AutoChipAH}, and patch bugs in the design based on precomputed or LLM-generated testbenches \cite{xu2024meic, qayyum2024bugs, Tsai2023RTLFixerAF, Huang2024TowardsLV, Yao2024HDLdebuggerSH}. Moreover, there has also been work in not just utilizing existing LLMs for verification, but also to fine-tune open-source models to better equip them with hardware specific domain knowledge. ChipNeMo \cite{Liu2023ChipNeMoDL} uses various domain-adaptation strategies including adaptive tokenization, pretraining, retrieval, and so on. \cite{Fu2023LLM4SecHWLD} utilize version-control data to compile a dataset of open-source hardware design bugs and corresponding fixes. 

These works confirm that while LLMs can generate hardware artifacts, domain-specific validation and feedback remain critical. Overall, these systems generally treat the LLM as a one-shot generator, not as an interactive agent capable of experimentation. Moreover, obtaining a deep understanding of the DUT is crucial for reliable verification, but limitations including limited-context windows of LLMs have prevented design-understanding techniques that can scale up with the complexity of designs \cite{Hassan2024LLMGuidedFV, Qayyum2024LateBR}. Approaches such as VerilogReader \cite{Ma2024VerilogReaderLH} use LLMs to enhance DUT code with natural language descriptions to provide greater context of intended functionalities in downstream LLM-tasks. Similarly, RTLExplain \cite{Chi2025RTLExplainAS} generates top-level module summaries via a bottom-up, data-dependency-aware method: extracting a module-dependency graph, producing summaries in reversed topological order, and compiling them into a project-level knowledge base. This achieved 37\% accuracy with a RAG agent, compared to 27\% with conventional RAG. Once again, the design-understanding obtained through such methods is restricted by their exclusive dependence on static RTL and design specifications, whereas \sys takes advantage of powerful hardware-specific tools that allow AI agents to iteratively verify and refine their design-understanding.

%% file: text/conclusion.tex
\section{Conclusion}
\label{sec:conclusion}

We present \sys{}, 
  a general methodology to develop comprehensive
  design understanding capabilities
  through iterative experimentation. 
We present a generally-applicable
  implementation of \sys{}
  which is useful in many potential 
  AI-assisted flows.
We demonstrate how
  experimentation with \sys
  produces deeper design understanding
  that has concrete, positive effects on 
  downstream tasks---in our case, formal verification.